\documentclass{aastex61}

\usepackage{graphicx}	
\usepackage{amssymb}
\usepackage{amsmath}
\usepackage{natbib}
\usepackage{color}
\usepackage{url}
\usepackage{placeins}
\usepackage{float}
\usepackage{color}
\usepackage{rotating}

\newcommand\aastex{AAS\TeX}

\newcommand{\ie}{{\it i.e. }}
\newcommand{\eg}{{\it e.g.\,}}
\newcommand{\cf}{{\it cf. }}
\newcommand{\insitu}{{\it in situ }}
\newcommand{\kmps}{{$\mathrm{km~s}^{-1}$}}
\newcommand{\gammaunit}{{$\cdot10^{-7} ~\mathrm{km}^{-1}$}}
\DeclareMathOperator\erf{erf}

\submitjournal{ApJ}

\shorttitle{\aastex\ Drag based ensemble model}
\shortauthors{Dumbovi\'{c} et al.}

\begin{document}

\title{Drag-Based Ensemble Model (DBEM) for Coronal Mass Ejection Propagation}

\correspondingauthor{Mateja Dumbovi\'{c}}
\email{mateja.dumbovic@uni-graz.at}

\author[0000-0002-8680-8267]{Mateja Dumbovi\'{c}}
\affil{Institute of Physics, University of Graz, Universit\"atsplatz 5, A-8010 Graz, Austria}

\author[0000-0002-4066-726X]{Ja\v{s}a \v{C}alogovi\'{c}}
\affiliation{Hvar Observatory, Faculty of Geodesy, University of Zagreb, Ka\v{c}i\'{c}eva 26, HR-10000, Zagreb, Croatia}

\author[0000-0002-0248-4681]{Bojan Vr\v{s}nak}
\affiliation{Hvar Observatory, Faculty of Geodesy, University of Zagreb, Ka\v{c}i\'{c}eva 26, HR-10000, Zagreb, Croatia}

\author[0000-0003-4867-7558]{Manuela Temmer}
\affil{Institute of Physics, University of Graz, Universit\"atsplatz 5, A-8010 Graz, Austria}

\author[0000-0001-9177-8405]{M. Leila Mays}
\affiliation{NASA Goddard Space Flight Center, Greenbelt, MD 20771, USA}

\author[0000-0003-2073-002X]{Astrid Veronig}
\affil{Institute of Physics, University of Graz, Universit\"atsplatz 5, A-8010 Graz, Austria}
\affil{Kanzelh\"ohe Observatory for Solar and Environmental Research, University of Graz, Universit\"atsplatz 5, A-8010 Graz, Austria}

\author[0000-0001-9124-6644]{Isabell Piantschitsch}
\affil{Institute of Physics, University of Graz, Universit\"atsplatz 5, A-8010 Graz, Austria}

\begin{abstract}

The drag-based model (DBM) for heliospheric propagation of coronal mass ejections (CMEs) is a widely used analytical model which can predict CME arrival time and speed at a given heliospheric location. It is based on the assumption that the propagation of CMEs in interplanetary space is solely under the influence of magnetohydrodynamical drag, where CME propagation is determined based on CME initial properties as well as the properties of the ambient solar wind. We present an upgraded version, covering ensemble modelling to produce a distribution of possible ICME arrival times and speeds, the drag-based ensemble model (DBEM). Multiple runs using uncertainty ranges for the input values can be performed in almost real-time, within a few minutes. This allows us to define the most likely ICME arrival times and speeds, quantify prediction uncertainties and determine forecast confidence. The performance of the DBEM is evaluated and compared to that of ensemble WSA-ENLIL+Cone model (ENLIL) using the same sample of events. It is found that the mean error is $ME=-9.7$ hours, mean absolute error $MAE=14.3$ hours and root mean square error $RMSE=16.7$ hours, which is somewhat higher than, but comparable to ENLIL errors ($ME=-6.1$ hours, $MAE=12.8$ hours and $RMSE=14.4$ hours). Overall, DBEM and ENLIL show a similar performance. Furthermore, we find that in both models fast CMEs are predicted to arrive earlier than observed, most probably owing to the physical limitations of models, but possibly also related to an overestimation of the CME initial speed for fast CMEs.

\end{abstract}


\section{Introduction}
\label{intro}

As coronal mass ejections (CMEs) are dominating space weather effects, including potentially harmful impacts for Earth, the forecasting of CMEs is one of the major challenges for space weather forecast. Therefore, in recent years many CME and their associated shock propagation models have been developed by research groups around the globe and space weather forecast centers regularly implement some of these propagation models to alert about a possible arrival of potentially threatening CMEs (\eg Space Weather Prediction Center of the National Oceanic and Atmospheric Administration, SWPC/NOAA, in USA, Met Office in UK or Solar Influences Data Analysis Center, SIDC, in Belgium).

The propagation models differ based on the input, approach, assumptions and complexity \citep[for overview see \eg][and references therein]{zhao14} and vary from simple empirical models \citep[\eg][]{gopalswamy01}, neural network models \citep[\eg][]{sudar16}, analytical drag-based models with different geometries \citep[\eg][]{vrsnak14,rollett16}, various kinematic shock propagation models \citep[\eg][]{dryer01,zhao16,takahashi17} to complex numerical 3D MHD models such as the H3DMHD model \citep{wu11} or WSA-ENLIL+Cone model \citep{odstrcil04}. Despite these differences, in general most of the models show a surprisingly comparable performance, where the prediction errors are mostly found within the 24-hour interval and the mean absolute error is $\sim10$ hours \citep[\eg][]{gopalswamy01,li08,vrsnak14,mays15,sudar16,wold17}.

The drag-based models are based on the concept of MHD drag, which, unlike the kinetic drag effect in a fluid, is presumed to be caused primarily by the emission of MHD waves in the collisionless solar wind environment and acts to adjust the CME speed to the ambient solar wind \citep{cargill96}. The concept of drag relies on the observational fact that slow CMEs accelerate whereas fast CMEs decelerate \citep[\eg][]{sheeley99, gopalswamy00} and is supported by numerous studies \citep[\eg][and references therein]{vrsnak07,temmer11,liu13,hess14,sachdeva15}. \citet{vrsnak07} proposed that the equation describing the aerodynamic drag can be utilised to establish a simple kinematical drag-based model (DBM) for CME propagation, which was since developed by implementing the cone geometry of a CME and performing parametric analysis to empirically determine the drag parameter $\gamma$ \citep{vrsnak13,vrsnak14,zic15}. The DBM is available at Hvar Observatory website as an online tool\footnote{\url{http://oh.geof.unizg.hr/DBM/dbm.php}}, it is one of the European Space Agency (ESA) space situational awareness (SSA) products\footnote{\url{http://swe.ssa.esa.int/heliospheric-weather}}, is one of the models available at the Community Coordinated Modeling Center (CCMC)\footnote{\url{https://ccmc.gsfc.nasa.gov}}, and is incorporated into the automatic COMESEP system \citep{crosby12,dumbovic17}. The performance of DBM was shown to be comparable to that of other propagation models \citep{vrsnak14} and was furthermore found to agree well with WSA-ENLIL+Cone model, which is one of the most extensively used CME propagation models in space weather operations world-wide.

The resemblance in the performance of very different propagation models indicates that the major drawback in more accurately forecasting CME arrival times and impact speeds is the lack of reliable observation-based input. In order to take into account the errors and uncertainties in the CME measurements that are used as model input and for quantifying the resulting uncertainties in the model predictions, ensemble forecasting is widely used. Recently, \citet{mays15} used an ensemble modelling approach to evaluate the sensitivity of WSA-ENLIL+Cone model (hereafter ENLIL) simulations to initial CME parameters and provide a probabilistic forecasting of CME arrival time. Ensemble modelling takes into account the variability of observation-based model input by making an ensemble, \ie sets of $n$ CME observations to calculate a distribution of predictions and forecast the confidence in the likelihood of the CME arrival. We use the ensemble approach in DBM and present the newly developed DBEM. The model is evaluated on the same data set as \citet{mays15} to be compared to ENLIL ensemble results.

\section{Data and method}
\label{data}

DBEM is based on DBM with assumed cone geometry for the CME, where the leading edge is initially a semicircle, spanning over the full angular width of the CME and flattens as it evolves in time \citep[described in][]{zic15}. The model assumes a constant solar wind speed, $w$, and drag parameter, $\gamma$, which is in general valid for distances beyond $R>15\mathrm{R_{SUN}}$, where the CME moves in an isotropic solar wind spreading out at a constant speed and the fall-off of the ambient density is at the same rate as the CME expansion \citep[see][]{vrsnak13,zic15}. As this assumption is clearly not valid for CME-CME interaction events \citep{temmer12}, we do not consider such interaction events in our study. We note that the constant $w$ and $\gamma$ assumption can be generally valid even for CMEs moving in high speed streams, assuming they encounter high speed streams relatively close to the Sun.

The input parameters derived from observations and used as input for the DBEM are the CME speed, halfwidth, and propagation direction (longitude) defined at a certain distance/time from the Sun. Since this is a 2D model that operates in the ecliptic plane DBEM does not use CME latitude as an input. The solar wind speed (the radial component in the ecliptic plane) and the drag parameter, $\gamma$, complete the input values. In a first step, for a single CME, we use an ensemble of $n$ measurements of the same CME as \citet{mays15} (see Section \ref{CMEin}). Next, the variability of DBM parameters (solar wind speed, $w$ and drag parameter, $\gamma$) is taken into account, where $m$ synthetic values of both $w$ and $\gamma$ are produced (see Section \ref{w&gamma}). These synthetic values are combined with an ensemble of $n$ CME measurements, to give a final ensemble of $n\cdot m^2$ members as an input, which, after $n\cdot m^2$ runs, produces a distribution of $n\cdot m^2$ calculated CME transit times and arrival speeds.

\subsection{CME initial parameters}
\label{CMEin}

We use the sample compiled and analysed by \citet{mays15}, which consists of 35 CMEs and associated interplanetary CMEs (ICMEs, if detected at Earth) in the time period January 2013 to July 2014. All CME measurements and simulation summary results are available at \url{https://iswa.ccmc.gsfc.nasa.gov/ENSEMBLE/}. The CME initial parameters are determined using the Stereoscopic CME Analysis Tool (StereoCAT) developed by CCMC. StereoCAT tracks specific CME features, based on triangulation of transient CME features manually identified using two different coronagraph fields-of-view. From this the 3D speed and position is derived for a CME as well as its (projected) width. To gather an ensemble of measurements, the CME leading edge height was measured for two different times in each coronagraph image for two different coronagraph viewpoints and then the procedure was repeated $k$ times to obtain an ensemble of $n=k^2$ CME measurements \citep[for details see][]{mays15}. Each ensemble member has a specific set of initial CME parameters - speed, width, longitude and latitude. Given that a typical run of the whole ensemble using ENLIL simulations is 80--130 min (depending on the computing power), for each event an optimal spread of input parameters was selected ($n=12, 16, 24, 36, 48$). It should be noted that the input is more suitable for the 3D ENLIL WSA+cone model, which uses both longitude and latitude as positional parameters and 3D velocity as input speed, whereas for DBEM the radial component of velocity in the ecliptic plane would be a more suitable input.

We refine the sample compiled by \citet{mays15} excluding events with CME-CME interaction and events where \insitu arrival times could not be determined exactly and unambiguously. The resulting sample consists of 25 CMEs and for each event we use as ensemble model input the derived speed, halfwidth, longitude, and start time. The start time corresponds to the starting distance, which is restrained by the ENLIL inner boundary corresponding to $R=21.5\mathrm{R_{SUN}}$ and is also suitable for DBM due to the assumption of constant $w$ and $\gamma$ and furthermore because after that distance the drag is the dominant force governing the propagation of ICMEs for a large subset of CMEs \citep{sachdeva15,sachdeva17}.

\subsection{Solar wind speed and drag parameter $\gamma$}
\label{w&gamma}

\begin{figure*}
\plotone{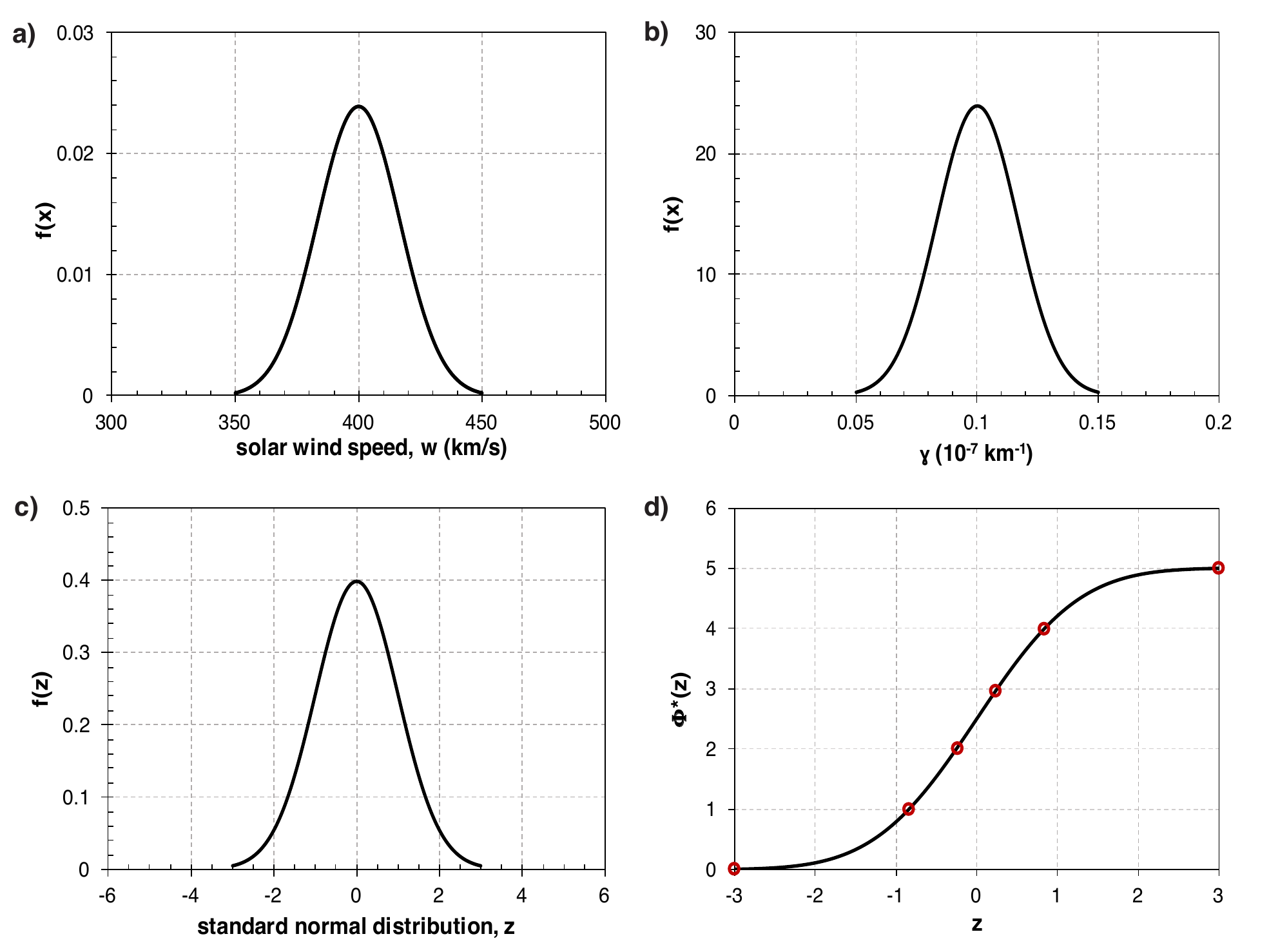}
\caption{Normal distributions for a) $w=(350\pm50)$ \kmps and b) $\gamma=(0.1\pm0.05)$ \gammaunit; c) standard normal distribution obtained with substitution $z=(x-\mu)/\sigma$; d) cumulative standard normal distribution normalised so that it is defined on an interval [0,$m-1$], where $m=6$ is the number of synthetic values in this example ($\Phi_{i} ^*(z)=0, 1, 2, 3, 4, 5$ are outlined in red).
\label{fig1}}
\end{figure*}

As input values for the solar wind speed, $w$, and drag parameter, $\gamma$, we do not use measured values, but empirical values derived in previous studies by \citet{vrsnak13, vrsnak14}. Therefore, to include their uncertainty in DBEM similarly to the uncertainty of the CME input (CME ensembles), synthetic values are needed. Synthetic values of $w$ and $\gamma$ for a specific event can be produced assuming that the real measurements of these parameters follow a normal distribution fully defined by the following expression:

\begin{equation}
x=\bar{x}\pm\Delta x\,,
\label{eq1}
\end{equation}

\noindent where $\bar{x} =\mu$ is the mean of the normal distribution and $\Delta x=3\cdot\sigma$ defines a range where 99,7\% measurements are found ($\sigma$ is the standard deviation). The normal distribution for a random variable $x$, $f(x)$, is defined for a specific $\mu$ and $\sigma$, where $w$ and $\gamma$ are treated as random variables described by a normal distribution. In Figures \ref{fig1}a and \ref{fig1}b examples of normal distributions are shown for $w=(350\pm50)$ \kmps and $\gamma=(0.1\pm0.05)$ \gammaunit, respectively. Based on our assumption, these are the distributions we would obtain by making real measurements of $w$ and $\gamma$, where their real values are found in the intervals $w=(350\pm50)$ \kmps and $\gamma=(0.1\pm0.05)$ \gammaunit, respectively. 

Using the substitution $z=(x-\mu)/\sigma$, a corresponding standard normal distribution (SND), $f(z)$, is obtained, which is normalised with respect to the original distribution so that $\mu=0$ and $\sigma=1$ (shown in Figure \ref{fig1}c). The cumulative SND, which is the probability that $z$ will take a value less than or equal to some $z_0$ (\ie gives the area under $f(z)$ from $-\infty$ to $z_0$), can be written as:

\begin{equation}
\Phi(z)=\frac{1}{2}\bigg(1+\erf{(\frac{z}{\sqrt{2}})}\bigg)\,,
\label{eq2}
\end{equation}

\noindent where the $\erf{(z)}$ is the Gauss error function defined as:

\begin{equation}
\erf{(z)}=\frac{2}{\sqrt{\pi}}\int_{0}^{z_0} e^{-z^2}dz\,.
\label{eq3}
\end{equation}

$\Phi(z)$ is a continuous function defined on an interval [0,1]. However, if multiplied with a normalisation factor $m-1$, where $m$ is the number of synthetic values we wish to produce, we obtain a new continuous function $\Phi^*(z)$ defined on an interval [0,$m-1$], which can be used to obtain $m$ different values of $z$:

\begin{equation}
z_i=-\sqrt{2}\cdot \erf^{-1}{\bigg(1-2\frac{\Phi_{i}^*(z)}{m-1}\bigg)},  \qquad \Phi_{i} ^*(z)=0,1,2,...,m-2,m-1\,.
\label{eq4}
\end{equation}

\begin{figure}
\plotone{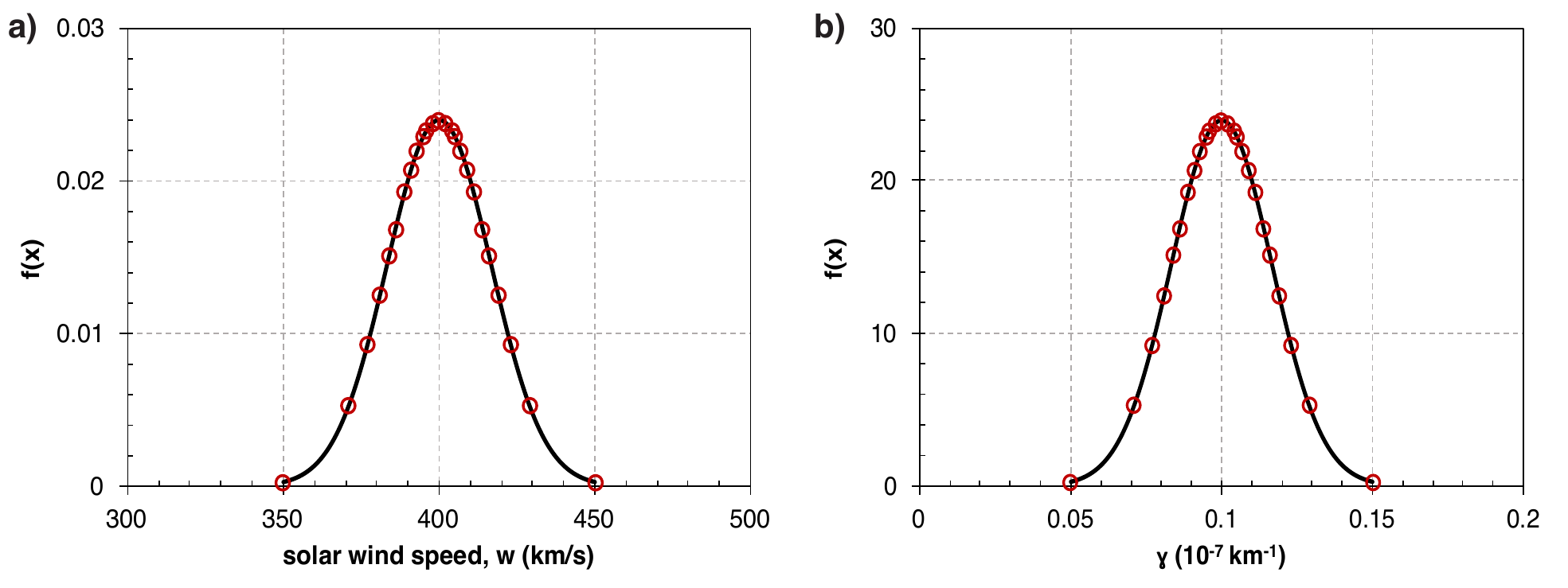}
\caption{Normal distributions for a) $w=(350\pm50)$ \kmps and b) $\gamma=(0.1\pm0.05)$ \gammaunit and corresponding synthetic values (red circles).
\label{fig2}}
\end{figure}

This is graphically represented in Figure \ref{fig1}d on an example where $m=6$. Each $z_i$ corresponds to a certain $x_i$ based on the substitution $z_i=(x_i-\mu)/\sigma$. Therefore, for a given number $m$ and a parameter defined according to Equation \ref{eq1}, $m$ synthetic values of the parameter can be produced based on Equation \ref{eq4}. An example is given in Figures \ref{fig2}a and  \ref{fig2}b, where $m=25$ synthetic values are shown against the normal distributions for $w=(350\pm50)$ \kmps and $\gamma=(0.1\pm0.05)$ \gammaunit, respectively.

\begin{figure*}
\plotone{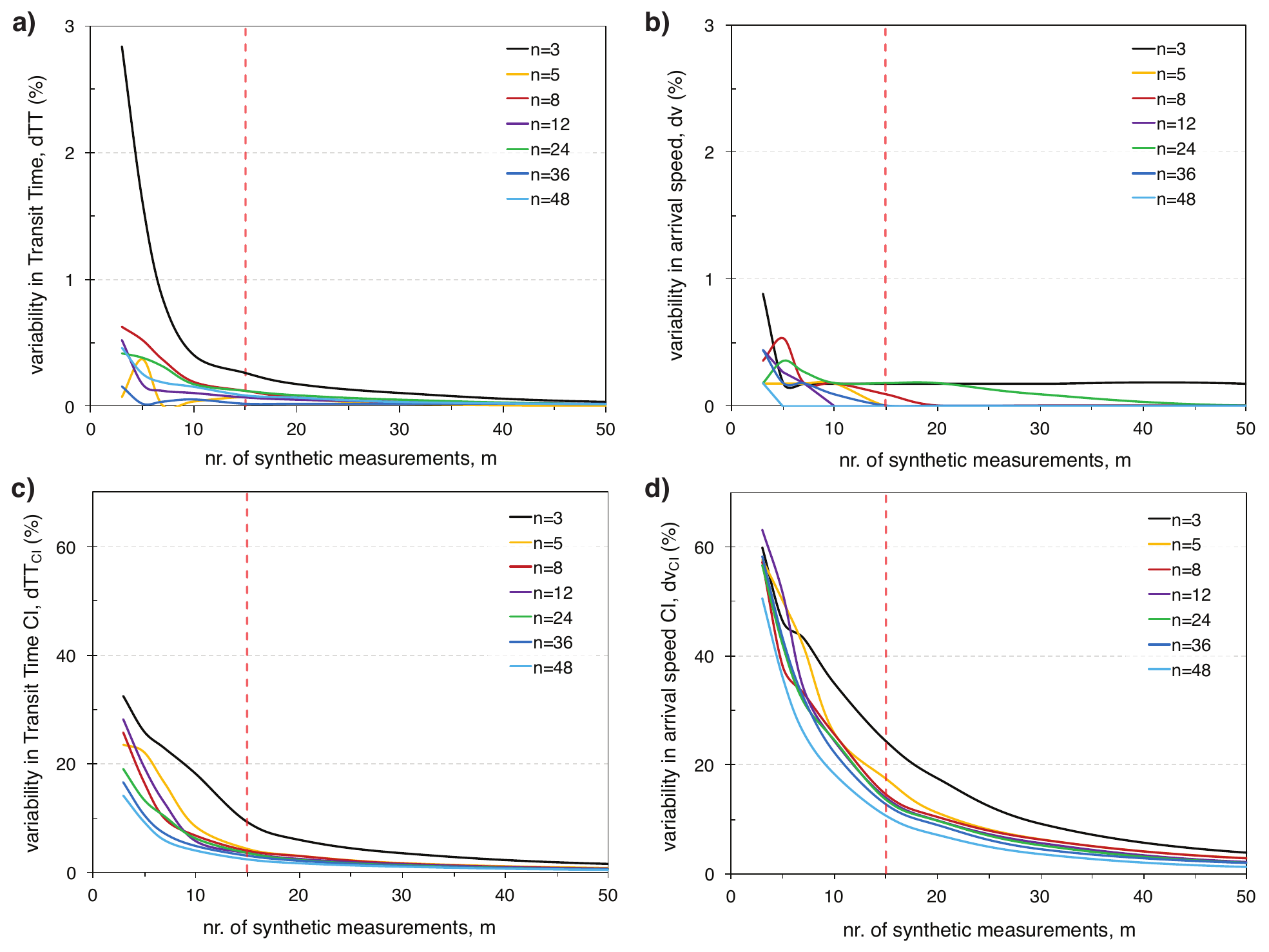}
\caption{The variability of the distribution median for a) CME transit time and b) arrival speed, and of the $95\%$ confidence interval for c) CME transit time and d) arrival speed depending on the number of CME ensemble members, $n$ and number of solar wind speed and $\gamma$ parameter synthetic values, $m$. The variabilities are normalised to values corresponding to $m=100$ (for each CME ensemble separately). The red dashed line marks the optimal value $m=15$ (for details see main text).
\label{fig3}}
\end{figure*}

Based on Figure \ref{fig2} it is obvious that the synthetic values will reflect the normal distribution closer if larger $m$ is chosen. However, very large $m$ will increase the number of runs and consequently computing time. Therefore, the optimum $m$ needs to be determined. For that purpose, we randomly selected a CME from our list which has a total of 48 ensemble members (2014 February 12). We then randomly selected 3, 5, 8, 12, 24, and 36 ensemble members to simulate respective ensemble sizes of the same CME (different ensemble sizes are denoted with $n$). For each of the ensembles we made model runs using different number of synthetic values $m=3, 5, 7, 10, 15, 20, 30, 50, 100$ and derived distributions of transit times, $TT$(h), and arrival speeds, $v$(\kmps), from which we calculate the median and $95\%$ confidence interval. The model runs include both the variability of the CME input ($n$) and the variability of the model parameters ($m$). In order to observe how the values of distribution median and $95\%$ confidence interval change with $n$ and $m$, we normalised for each $n$ the difference of the $m$-th median to the median corresponding to a value of 100 in the following way:

\begin{equation}
dX (n,m)= \frac{|X(n,m)-X(n,100)|}{X(n,100)}\,,
\label{eq5}
\end{equation}

\noindent where $X(n,m)$ is the median of $TT$ and $v$ for CME transit time and arrival speed, respectively. We applied the same equation to $95\%$ confidence interval as well, where for each $n$ the difference of the $m$-th $95\%$ confidence interval is normalised to the $95\%$ confidence interval corresponding to $m=100$ according to Equation \ref{eq5} with $X(n,m)$ being $TT_\mathrm{range}$ and $v_\mathrm{range}$ for CME transit time and arrival speed, respectively. This is shown in Figure \ref{fig3} where it can be seen that the variability of the median for both arrival speed and transit time is quite small ($<1\%$ for arrival speed, $<3\%$ for transit time) and decreases with both $n$ and $m$. The variability of the $95\%$ confidence interval is much larger (can go up to $30\%$ for transit time and even $60\%$ for arrival speed), but also decreases quickly with increasing $n$ and $m$. Both the median and the $95\%$ confidence interval converge quite fast towards the value corresponding to $m=100$. At $m=15$ the variability of the median is already below $0.5\%$ for all CME ensemble sizes $n$ for both $TT$ and $v$, whereas the variability of the $95\%$ confidence interval is below $5\%$ for $TT$ and around $10\%$ for $v$. We note that the typical prediction errors of the CME transit time are $\sim10$ hours (as described in Section \ref{intro}), thus a $5\%$ variation in a confidence interval would correspond to less than $\sim1$ hour and a $0.5\%$ variation of the median would be of the order of magnitude of $\sim1$ minute. Therefore, as an optimal value we choose $m=15$. For an ensemble of $n=48$ CME inputs, with this optimal value of $m=15$ synthetic $w$ and $\gamma$ values the total number of model runs is $n\cdot m^2=10800$, resulting in a computational time of several minutes on an average PC.

For simplicity we use the same value of mean solar wind speed and $\gamma$ parameter for all events in the sample. Due to the weak solar activity throughout the last solar cycle, which was reflected on the solar wind and interplanetary magnetic field \citep[\eg][]{mccomas13}, we use $w=(350\pm50)$ \kmps. DBM tracks the leading edge of the ICME ejecta, whereas ENLIL tracks the shock front. However, it was shown that there is in general a good agreement between the two with a convenient selection of the $\gamma$ parameter \citep[$\gamma=0.1$\gammaunit, see][]{vrsnak14}. Therefore, in order to estimate shock arrival with DBEM we select $\gamma=(0.1\pm0.05)$ \gammaunit. With these values of $w$ and $\gamma$ we determine $m=15$ synthetic values of $w$ and $\gamma$ with the procedure described above. It should be noted that the simplification of using average solar wind conditions results in less realistic solar wind background than the one used by ENLIL.

\section{Results and Discussion}
\label{results}	

\begin{figure}[ht!]
\plotone{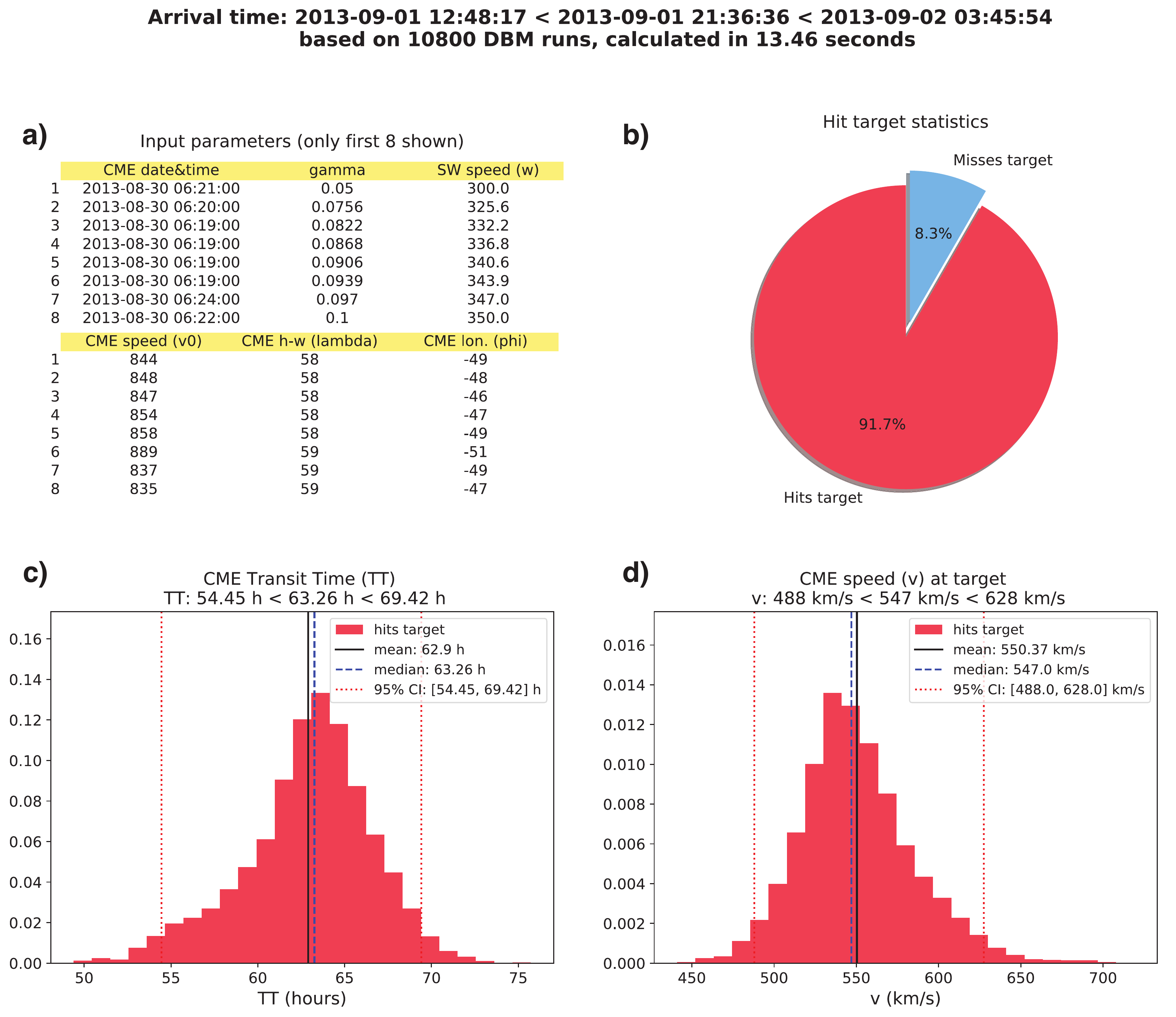}
\caption{DBEM results for the 2013 August 30 CME: a) a quicklook table of the (partial) input; b) a pie chart with the probability of arrival; c) the distribution of the CME transit time; d) the distribution of the arrival speed. Distributions of the CME transit time and arrival speed are calculated only for ensemble members that hit Earth and are given with the corresponding distribution parameters (mean, median and $95 \%$ confidence interval). The median (most likely) arrival time and the $95 \%$ confidence interval are given on top with a basic performance record (number of runs and calculation time).
\label{fig4}}
\end{figure}

For each ensemble member, \ie individual run, DBEM calculates whether or not the CME will hit or miss the Earth. The corresponding condition can be written as $\omega \geq |\phi|$, where $\omega$ is the CME halfwidth and $\phi$ is the CME source position longitude. For the whole ensemble DBEM calculates the probability of the arrival as $p=n_\mathrm{hits}/n_\mathrm{tot}$, where $n_\mathrm{hits}$ is the number of ensemble members that are calculated to hit Earth and $n_\mathrm{tot}$ is the total number of all ensemble members. The probability of arrival for a CME is displayed as a pie chart, as shown in the upper right panel of Figure \ref{fig4} where DBEM results for the example CME 2013 August 30 are given. As can be seen in Figure \ref{fig4} the probability of arrival for this event, as calculated by DBEM is 0.92 (91.7\%, red part of the pie chart). The CME input for this event consists of 48 ensemble members with start time range 05:59 UT to 06:30 UT, CME speed range $810-1012$ \kmps, longitude range E55$^{\circ}-$E35$^{\circ}$, and halfwidth range 41$^{\circ}-63^{\circ}$. This is supplemented with 15 synthetic values of solar wind speed in the range $300-400$ \kmps and 15 synthetic values of the drag parameter $\gamma$ in the range $0.05-0.15$ \gammaunit. Therefore, the size of the whole ensemble for this example is 10800, \ie the results are based on 10800 DBM runs. A table of the first eight ensemble members (input lines) is given in the top left panel as a quicklook visualisation of the input. 

The bottom panels in Figure \ref{fig4} show transit time, $TT$, and arrival speed, $v$, distributions, calculated only based on the runs for ensemble members calculated to hit Earth (the red part of the pie chart). The expected range is given by the $95\%$ confidence interval. The most likely arrival date and time and corresponding range are given on the top of Figure \ref{fig4} and are also calculated only based on ensemble members predicted to hit Earth. The output also provides a quick performance record - the number of runs and the computing time. The example event shown in Figure \ref{fig4} was run on an average PC. Depending on the number of CPUs and their speed, the current DBEM version can make thousand (on single thread/CPU) or more runs per second and can run the whole ensemble in several minutes, which is very fast compared to ENLIL runtime ($>1$ hour on a high-performance machine, see Section \ref{CMEin}). We perform the DBEM forecast evaluation for a set of 25 events, and when possible, compare the outcome to that of ENLIL. Forecast evaluation is based on DBEM and ENLIL calculations of $TT$, DBEM calculation of $v$, and \insitu observations of $TT$ and $v$ presented in Table \ref{tab1} \citep[for additional information on the median values of the CME input we refer the reader to Table 1 of][]{mays15}. We use the peak value of the \insitu speed to be consistent in all events, due to the fact that sometimes only shock/sheath is encountered, sometimes only ICME magnetic structure and sometimes both.

\begin{sidewaystable}
\centering
\begin{tabular}{cccccccccccccccccc}
\hline
CME&	ICME&	Observed&	Observed&	\multicolumn{9}{c}{\textbf{DBEM}}&		\multicolumn{5}{c}{\textbf{ENLIL}}\\
start&	arrival&	$TT$&	$v$&		$p_i$&	$TT$&		$\Delta T_{\mathrm{err}}$&		$TT_{\mathrm{min}}$&	$TT_{\mathrm{max}}$&	$v$&		$\Delta v_{\mathrm{err}}$&	$v_{\mathrm{min}}$&
$v_{\mathrm{max}}$&	$p_i$&	$TT$&		$\Delta T_{\mathrm{err}}$&		$TT_{\mathrm{min}}$&	$TT_{\mathrm{max}}$\\

date&		date \& time&	(h)&		\kmps&	(\%)&	(h)&		(h)&		(h)&	(h)&	\kmps&	\kmps&	\kmps&	\kmps&	(\%)&	(h)&		(h)&		(h)&	(h)\\
\hline
\textbf{2013} & \multicolumn{17}{c}{}\\
\hline
04/11&	04/13 22:13&	62.8	&520			&100		&48.2	&-14.6	&40.4	&56.3	&653		&133		&556		&807			&100		&46.8	&-16		&41.4	&52.9\\
06/21&	06/23 03:51&	48.7	&720			&97.9	&34.8	&-13.9	&28.5	&42		&841		&121		&676		&1137		&97.9	&33.9	&-14.8	&30.3	&39.6\\
08/02&	--		  &	--	&--			&0		&--		&CR		&--		&--		&--		&--		&--		&--			&0		&--		&CR		&--		&--\\
08/07&	--		  &	--	&--			&100		&74		&FA		&--		&--		&--		&--		&--		&--			&100		&--		&FA		&--		&--\\
08/30&	09/02 01:56&	71.1	&510			&91.7	&63.3	&-7.8	&54.5	&69.4	&547		&37		&488		&628			&95.8	&53.7	&-17.4	&48.3	&58.3\\
09/19&	--		  &	--	&--			&0		&--		&CR		&--		&--		&--		&--		&--		&--			&0		&--		&CR		&--		&--\\
09/29&	10/02 01:15&	52.6	&630			&100		&50.8	&-1.8	&39.7	&62.4	&627		&-3		&526		&793			&100		&55.5	&2.9		&44.7	&64.6\\
10/06&	10/08 19:40&	53	&650			&100		&58.3	&5.3		&47.5	&64.6	&569		&-81		&503		&681			&91.7	&79.5	&26.5	&69.7	&89.6\\
10/22&	--		  &	--	&--			&100		&58		&FA		&--		&--		&--		&--		&--		&--			&95.7	&--		&FA		&--		&--\\
\hline
\textbf{2014} & \multicolumn{17}{c}{}\\
\hline
01/07&	01/09 19:39&	49.3	&480			&100		&26.3	&-23		&19.8	&33.9	&1019	&539		&772		&1474		&100		&29.9	&-19.4	&23		&39.1\\
01/30&	02/02 23:20&	78.9	&470			&87.5	&59		&-19.9	&44.6	&78.1	&563		&93		&459		&721			&54.2	&65.7	&-13.2	&53.4	&77.6\\
01/31&	--		  &	--	&--			&100		&68.7	&FA		&--		&--		&--		&--		&--		&--			&100		&--		&FA		&--		&--\\
02/12&	02/15 12:46&	79.1	&450			&100		&58.2	&-20.9	&49.7	&72.2	&562		&112		&475		&663			&100		&66.1	&-13		&57.4	&79.3\\
02/18&	02/20 02:42&	49.3	&690			&75		&59.5	&10.2	&30.3	&86.2	&558		&-132	&424		&926			&80.6	&63.1	&13.8	&34.4	&80.8\\
02/19&	02/23 06:09&	86.2	&510			&100		&53.1	&-33.1	&43.5	&62.2	&604		&94		&518		&743			&88.9	&68.4	&-17.8	&55.8	&81.6\\
02/25&	02/27 15:50&	62.7	&500			&54.2	&38.2	&-24.5	&29.5	&55.2	&763		&263		&580		&1038		&83.3	&45.1	&-17.6	&33.8	&65.8\\
03/23&	03/25 19:10&	63.4	&520			&43.8	&72		&8.6		&58.1	&85.5	&497		&-23		&431		&592			&79.2	&69.2	&5.8		&56.6	&81\\
03/23&	--		  &	--	&--			&8.3		&114.4	&CR		&--		&--		&--		&--		&--		&--			&0		&--		&CR		&--		&--\\
03/29&	--		  &	--	&--			&50		&75.2	&FA		&--		&--		&--		&--		&--		&--			&5.6		&--		&CR		&--		&--\\
04/02&	04/05 10:00&	68.4	&500			&31.2	&43.5	&-24.9	&37		&50.4	&710		&210		&598		&891			&87.5	&53.4	&-15		&43.3	&59.7\\
04/18&	04/20 10:20&	45.2	&700			&100		&37.8	&-7.4	&31.7	&43.8	&780		&80		&644		&1010		&100		&40		&-5.2	&36		&46.1\\
06/04&	06/07 16:12&	72.4	&610			&88.9	&79		&6.6		&66		&110.8	&460		&-150	&354		&539			&61.1	&77.1	&4.7		&69		&83.9\\
06/10&	--		  &	--	&--			&2.8		&61.4	&CR		&--		&--		&--		&--		&--		&--			&5.6		&--		&CR		&--		&--\\
06/19&	06/22 18:28&	73.3	&410			&100		&79.7	&6.4		&68		&92.3	&460		&50		&403		&530			&100		&71		&-2.3	&65.6	&78.6\\
06/30&	--		  &	--	&--			&0		&--		&CR		&--		&--		&--		&--		&--		&--			&0		&--		&CR		&--		&--\\
\hline
\end{tabular}
\caption{Observed and calculated transit times for the 25 CMEs under study using DBEM and ENLIL. $p_i$ is the probability of arrival, $TT$ is the transit time, and $v$ is the arrival speed. $\Delta T_{\mathrm{err}}=TT_{\mathrm{predicted}}-TT_{\mathrm{observed}}$ and $\Delta v_{\mathrm{err}}=v_{\mathrm{predicted}}-v_{\mathrm{observed}}$ are prediction errors, whereas $TT_{\mathrm{min}}$/$TT_{\mathrm{max}}$ and $v_{\mathrm{min}}$/$v_{\mathrm{max}}$ define the arrival spread (the 95\% confidence interval) for $TT$, and $v$, respectively. CR and FA denote correct rejection and false alarm, respectively (for explanation see main text).}
\label{tab1}
\end{sidewaystable}

The probability of arrival, $p_i$ is binned into a categorical yes/no forecast using a $p_i<15\%$ criterion for correct rejection, same as was used in \citet{mays15}. Thus a 2x2 contingency table for a binary event can be constructed and used to perform forecast evaluation \citep[see \eg][]{jolliffe03}, where CME arrival is regarded as an event, and the event forecast as well as the event observation can have two outcomes, yes or no. There are four possible combinations of forecast and observation outcomes: a ``hit" where the event was forecasted to hit Earth and observed, a ``miss" where the event was not forecasted to hit Earth but was observed, a ``false alarm" where the event was forecasted to hit Earth but was not observed and a ``correct rejection", where the event was neither forecasted nor observed. Following the procedure by \citet{mays15}, if $p_i<15\%$ and there were no \insitu signatures, the event (CME arrival) is considered as a ``correct rejection"; if $p_i>15\%$ and there were no \insitu signatures, the event is considered as a ``false alarm"; if $p_i>15\%$ and there are \insitu signatures, the event is considered as a ``hit"; and finally, if $p_i<15\%$ and there are \insitu signatures, the event is considered as a ``miss".

\begin{table}
\centering
\begin{tabular}{lccc}
\multicolumn{2}{c}{ }&	DBEM&	ENLIL\\
\multicolumn{4}{c}{ a) contingency table results}\\
\hline
Number of hits 				&$a$			&16	&16\\
Number of misses			&$c$			&0	&0\\
Number of false alarms		&$b$			&4	&3\\
Number of correct rejections 	&$d$			&5	&6\\
Number of events			&$N=a+b+d$	&25	&25\\
\hline
\multicolumn{4}{c}{}\\
\multicolumn{4}{c}{b) evaluation measures}\\
\hline
Correct rejection rate		&$d/(b+d)$ 					&55.6\%	&66.7\%\\
False alarm rate		& $b/(b+d)$ 					&44.4\%	&33.3\%\\
Correct alarm ratio		& $a/(a+b)$ 					&80.0\%	&84.2\%\\
False alarm ratio		& $b/(a+b)$ 					&20.0\%	&15.8\%\\
Brier score			& $BS$ (see Equation \ref{eq6}	)	&0.17	&0.18\\
\hline
\multicolumn{4}{c}{}\\
\multicolumn{4}{c}{c) prediction errors for $TT$ (h)}\\
\hline
\multicolumn{2}{l}{mean error ($ME$)}				&-9.7	&-6.1\\
\multicolumn{2}{l}{mean absolute error ($MAE$)}		&14.3	&12.8\\
\multicolumn{2}{l}{root mean square error ($RMSE$)}	&16.7	&14.4\\
\hline
\multicolumn{4}{c}{}\\
\multicolumn{4}{c}{d) prediction errors for $v$ (\kmps)}\\
\hline
\multicolumn{2}{l}{mean error ($ME$)}				&84		&--\\
\multicolumn{2}{l}{mean absolute error ($MAE$)}		&133		&--\\
\multicolumn{2}{l}{root mean square error ($RMSE$)}	&181		&--\\
\end{tabular}
\caption{Contingency table results, evaluation measures and prediction errors for DBEM and ENLIL (calculated based on the sample presented in Table \ref{tab1}).}
\label{tab2}
\end{table}

The number of hits, misses, false alarms and correct rejections for our sample is given in Table \ref{tab2}a for DBEM and ENLIL, and the outcome using different evaluation measures is given in Table \ref{tab2}b. Compared to ENLIL, DBEM has out of 25 events one false alarm more and one correct rejection less than ENLIL, which results in a somewhat lower performance. In the final row of Table \ref{tab2}b a Brier score ($BS$) is given, which quantifies the probability forecast errors:

\begin{equation}
BS=\frac{1}{N} \sum_{i=1}^{N}(p_i-o_i)^{2}\,,
\label{eq6}
\end{equation}

\noindent where $N$ is the total number of events, $p_i$ is the forecast probability that event $i$ will occur, and $o_i$ takes value 0 if event $i$ did not occur and 1 if the event $i$ did occur. For perfect forecast $BS=0$ and we can see that both DBEM and ENLIL are not too far away from this value.

In Table \ref{tab2}c transit time errors are given for DBEM and ENLIL, calculated for ``hit" events. It can be seen that DBEM has somewhat larger errors than ENLIL. One of the reasons is that the CME input used is more suitable for ENLIL than DBEM. Optimal CME input for ENLIL is a full 3D information of the CME as given in the sample (3D velocity, longitude and latitude) while for DBEM optimal input would be the radial velocity in the ecliptic plane. Possibly an even more important issue is the ambient solar wind state. ENLIL used a more realistic solar wind background, whereas DBEM assumed average solar wind conditions for all CMEs ($w\sim350$\,\kmps, $\gamma\sim0.1$\,\gammaunit) not taking into account possible event--to--event variability (\eg propagation through high speed streams). Another aspect which was not considered is a possible preconditioning effect, when an earlier CME ``clears the path" effectively reducing $\gamma$ \citep{liu14}, which can be even ten times lower than the present average value in extreme cases \citep{temmer15}. Therefore, it is reasonable to assume that DBEM would perform better if a more realistic solar wind background is considered and if the radial speed in the ecliptic plane was used. Nevertheless, overall these errors are still comparable to CME arrival time prediction errors reported in other studies \citep[see \eg][and references therein]{li08,vrsnak14, mays15,sudar16,wold17}.

\begin{figure}
\plotone{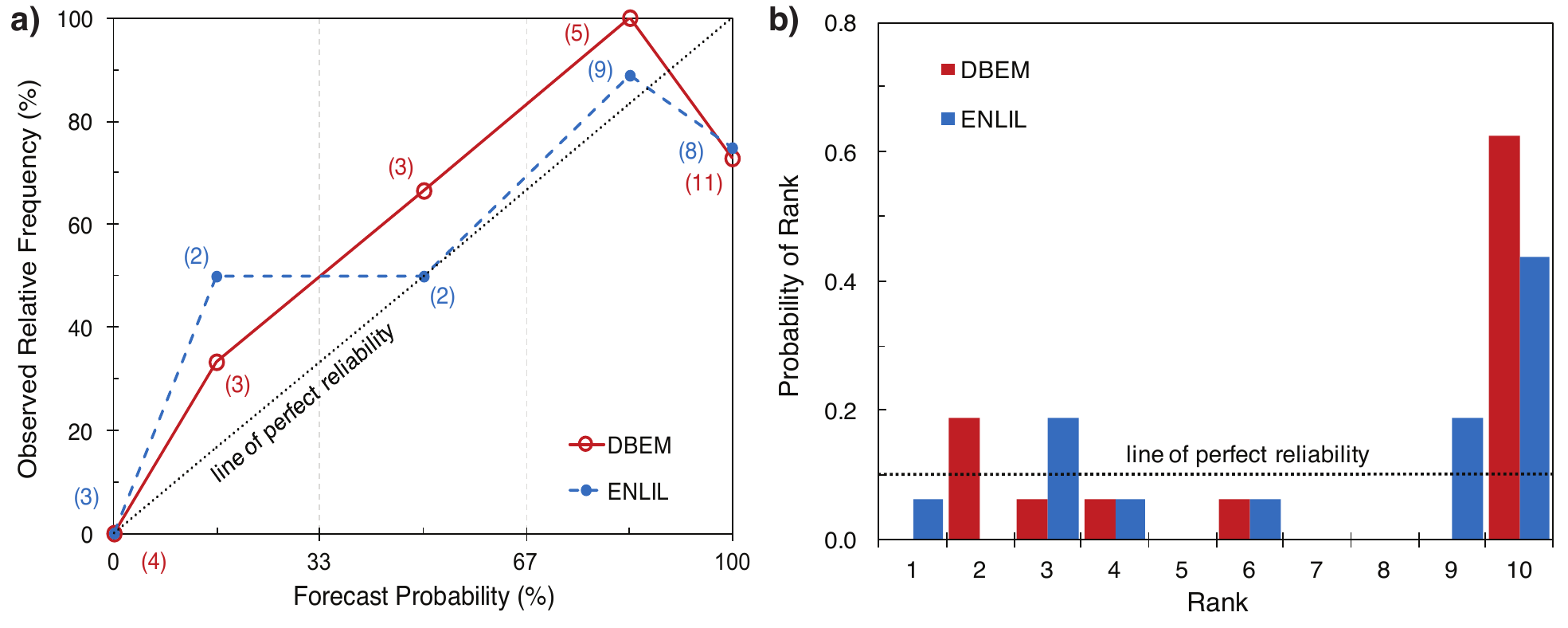}
\caption{a) Reliability diagram of the forecast probability of CME arrival for the whole sample of 25 events; b) Rank histogram for the 16 hits.
\label{fig5}}
\end{figure}

Next we test the performance of DBEM using the so called reliability diagram, which shows how well the predicted probabilities of an event correspond to their observed frequencies, \ie how well the model predicts the probability of arrival \citep[see \eg][]{jolliffe03}. In order to obtain the reliability diagram the events were binned according to their forecasted probability of arrival into 5 ``bins": $0\%$, [$0-33\%$], [$33-66\%$], [$66-100\%$], and $100\%$. For each bin an observed relative frequency was calculated as $\sum_{i=1}^{N_o}{O_i}/N_o$, where $N_o$ is the number of events corresponding to the bin and $O_i$ takes a value 0 if event $i$ did not occur and 100 if the event $i$ did occur. The line of perfect reliability is the identity line, \ie in a perfectly reliable forecast the forecast probability equals the observed relative frequency. The reliability diagram based on the selected sample for DBEM and ENLIL is shown in Figure \ref{fig5}a, where the number of events used in each calculation is shown next to each point. Although some points are calculated based on only 2 events observed, the diagram can reflect some general aspects of the DBEM and ENLIL forecast reliability. It can be seen that for the $100\%$ bin both DBEM and ENLIL overforecast, \ie predict higher probability of CME arrival than is observed, in agreement with a notable value of the false alarm rate in Table \ref{tab2}. For $0\%$ bin the point lies on the line of perfect reliability for both ENLIL and DBEM, however it should be noted that this is related to the fact that there are no missing alarms for neither model. With the intermediate bins one should be careful with drawing conclusions due to small number of calculations in some points, but at a descriptive level it would seem that both models slightly underforecast (ENLIL being closer to the line of perfect reliability). Similar conclusions were drawn for ENLIL by \citet{mays15} with an extended sample and slightly different selection of bins \citep[see Figure 9a in][]{mays15}.

In Figure \ref{fig5}b a rank diagram (also known as the ``Talagrand" diagram) is shown, which reflects how well the ensemble spread of the forecast represents the true variability of the observations, \ie whether observations statistically belong to the forecasted distributions \citep[see \eg][and references therein]{hamill01}. The rank diagram is constructed by sorting the $n$ ensemble members and the observation for each event from earliest to latest arrival time and ``counting" at which place we find the observation with respect to other ensemble members (denoted as ``rank"). For a perfect forecast, where we can statistically regard the observation as another member of the ensemble the observation is equally likely to occur in each of the $n+1$ possible ``ranks". Due to the fact that not all events have same ensemble sizes and moreover, that ensemble sizes for DBEM are drastically increased by introducing the $w$ and $\gamma$ synthetic values, we follow the same procedure as \citet{mays15} and normalise the rank number to $n=9$ (in total 10 possible ranks). The ``perfect reliability" line in Figure \ref{fig5}b represents an ideally flat distribution where each rank has the same probability (\ie 1.6 events per rank).

\begin{figure}
\plotone{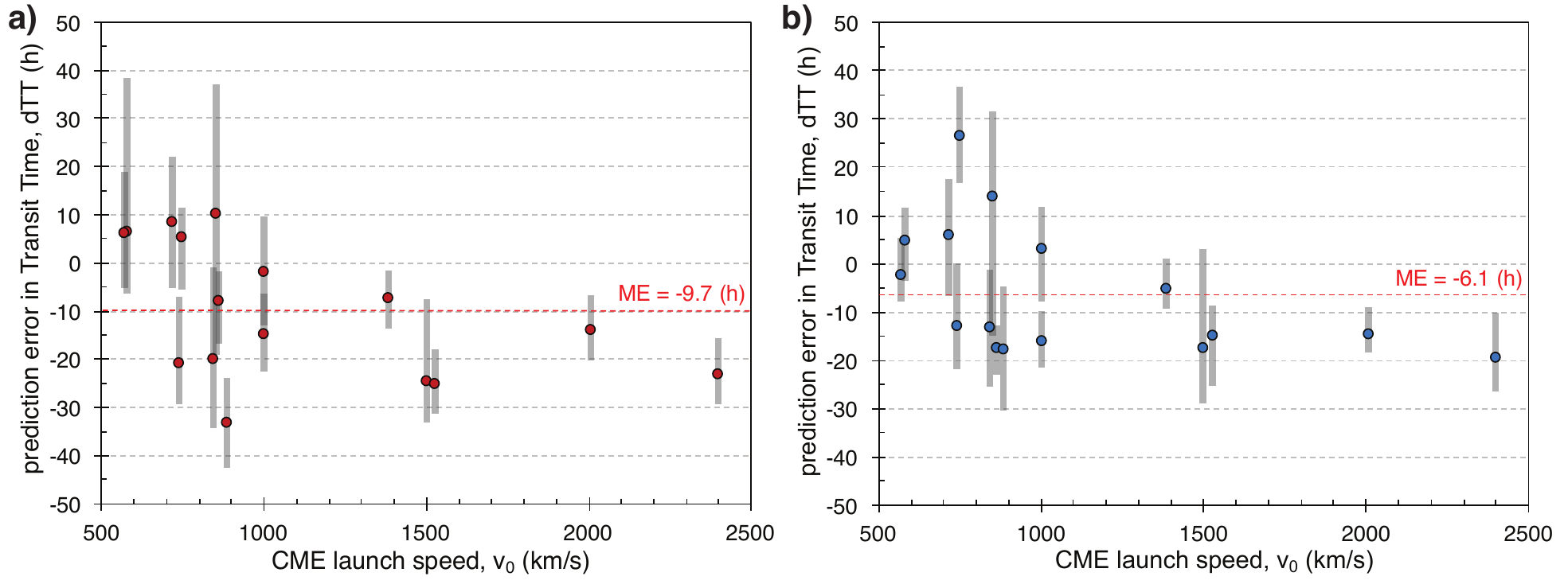}
\caption{CME arrival time prediction error plotted against the CME input speed for a) DBEM and b) ENLIL. Error bars represent the spread in the ensemble forecasts.
\label{fig6}}
\end{figure}

It can be seen, that the rank diagram for both ENLIL and DBEM deviates from the perfect reliability. For both models the rank diagram shows an U-shape which indicates under-variability of the ensemble and an asymmetrical shape, which indicates a bias. For both DBEM and ENLIL the possible bias is towards under-forecasting, \ie predicting smaller transit times than observed (also visible from the negative mean error presented in Table \ref{tab2}b). This is also supported by the fact that in only 38\% of events the observed $TT$ can be found within the DBEM prediction spread, whereas for ENLIL this percentage is somewhat higher (50\%). The possible reason for this bias could be related to fast CMEs. \citet{mays15} found that in their (extended) sample ENLIL generally predicted fast CMEs to arrive earlier than they were observed. To confirm this assumption, CME arrival time prediction errors are plotted against the CME input speed in Figure \ref{fig6}. An almost consistent negative prediction error can be seen for both DBEM and ENLIL for fast CMEs above $\sim$1000 \kmps, \cf Figure 8a by \citet{mays15}. This indicates that fast CMEs are indeed predicted to arrive earlier than observed for both ENLIL and DBEM, and results in a negative mean error and a bias towards under-forecasting. This might be related to model limitations, since they do not take into account all relevant physical processes. \citet{liu13,liu16} found that fast CMEs (with speed above 1000\,\kmps) differ from slower CMEs by a rapid deceleration process, which is not taken into account by DBEM and ENLIL. This would result in earlier predictions and overestimated arrival speed. Indeed, as seen in Table \ref{tab2}d and Figure \ref{fig8}a, DBEM has a tendency to overforecast the arrival speed for fast CMEs. It should also be noted that the role of CME-driven shocks is not considered in the drag based model \citep{reiner03,liu13,liu16}. DBM considers the physics of the magnetic structure of CMEs, not their related shocks, and it only estimates the shock arrival time using empirically obtained proxy value of $\gamma$ \citep{vrsnak14}. Another possible reason could be an overestimation of the CME initial speed for fast CMEs, as suggested by \citet{mays15}. Although DBEM and ENLIL use a range of CME initial speed as input, a large systematic overestimation of CME speed in the ensemble (\eg due to limited number of measurement points in fast events) could lead to the observed underforecast of transit times and overforecast of arrival speed.

\begin{figure}
\plotone{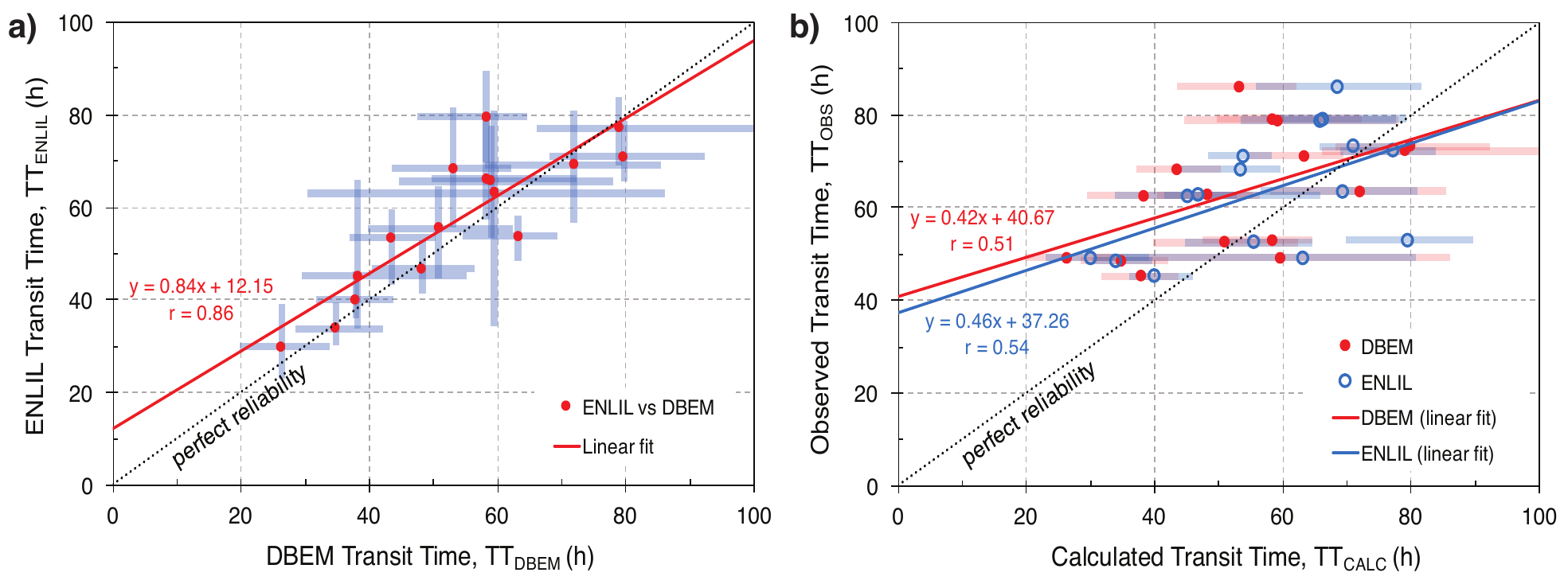}
\caption{a) ENLIL-calculated \textit{vs.} DBEM-calculated transit time;
b) Observed \textit{vs.} calculated transit time for ENLIL (blue) and DBEM (red).
Solid lines represent the linear best fits, with corresponding equations and correlation coefficients $r$. The dotted line represents the identity line (perfect match). Error bars represent the spread in the ensemble forecasts.
\label{fig7}}
\end{figure}

Finally, we examine the correlation between observed and calculated transit times. In Figure \ref{fig7}a ENLIL-calculated transit time is plotted as a function of DBEM-calculated transit time. It can be seen that there is some scatter and the linear best fit somewhat deviates from the identity line, but in general there is a good agreement between ENLIL-calculated and DBEM-calculated transit times, similar to the results obtained for non-probabilistic ENLIL and DBM by \citet{vrsnak14}. In Figure \ref{fig7}b, the observed transit time is plotted as a function of calculated transit time for ENLIL (blue) and DBEM (red). It can be seen that both for ENLIL and DBEM the linear best fit substantially deviates from the identity line. This is seen in the linear best fit coefficients, where the slope is much smaller than 1 and the intercept much larger than 0. Furthermore it can be seen that for higher calculated transit times the scatter is around the identity (perfect) line, whereas for smaller calculated transit times the scatter is above the identity line, which is related to the fact that fast CMEs are predicted to arrive earlier than observed. In Figure \ref{fig8}b, the observed arrival speed is plotted as a function of calculated arrival speed for DBEM. Similar to transit times, it can be seen that the linear best fit substantially deviates from the identity line, related to the overforecast of the arrival speed (for smaller values of calculated arrival speeds the scatter is around the identity line, whereas for larger values of calculated arrival speeds the scatter is below the identity line). We note that there seems to be an outlier with calculated $v=1019$\,\kmps. When the outlier is removed the slope of the best linear fit obtains a value of 0.36 and thus improves towards the identity line, and the correlation coefficient increases to $r=0.42$. We note that the outlier seems not to appear in the transit time calculation (Figure \ref{fig7}b), but only in the calculation of arrival speed (Figure \ref{fig8}b).

\begin{figure}
\plotone{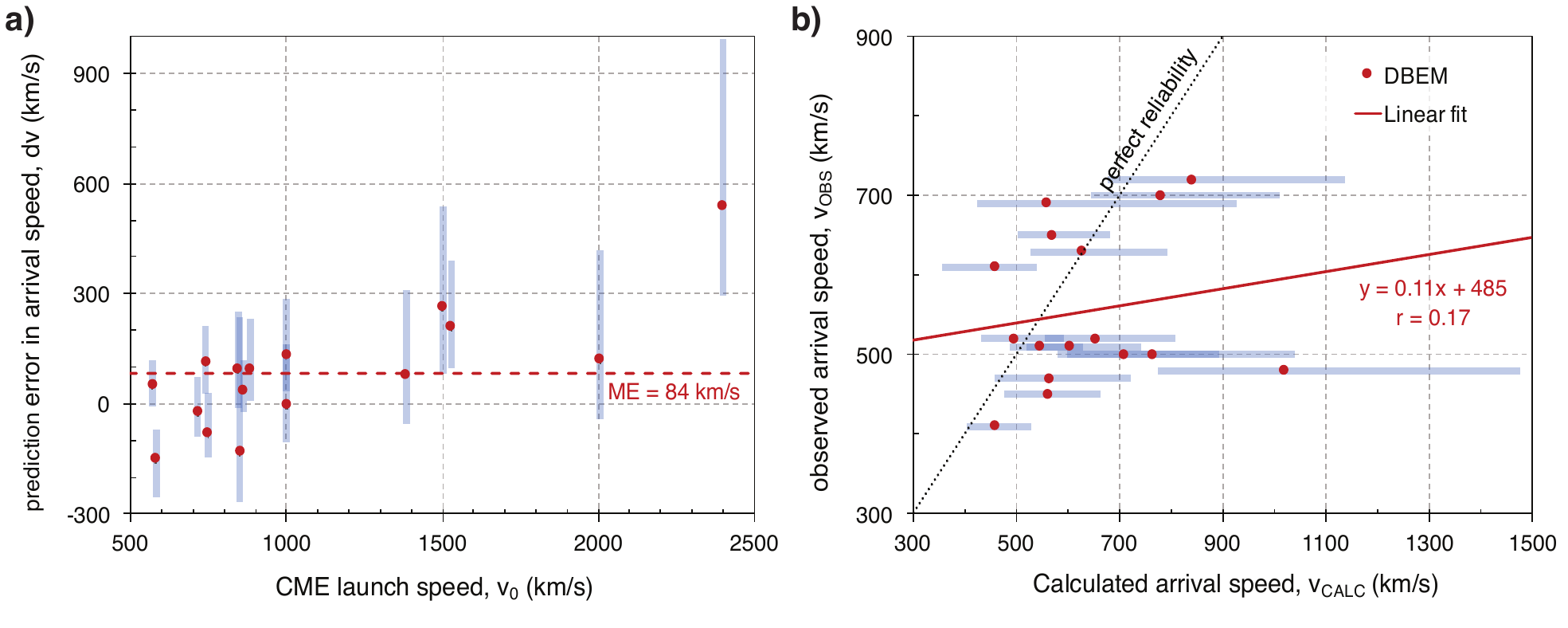}
\caption{a) CME arrival speed prediction error for DBEM plotted against the CME input speed;
b) Observed \textit{vs.} calculated arrival speed for DBEM.
Solid line represents the linear best fit, with corresponding equation and correlation coefficient $r$. The dotted line represents the identity line (perfect match). Error bars represent the spread in the ensemble forecasts.
\label{fig8}}
\end{figure}

\section{Summary and Conclusion}
\label{conclusion}

We present a probabilistic approach in the drag-based CME propagation modelling, the drag-based ensemble model (DBEM), which is available as an online tool at Hvar Observatory website\footnote{\url{http://oh.geof.unizg.hr}}, and as one of the European Space Agency (ESA) space situational awareness (SSA) products\footnote{\url{http://swe.ssa.esa.int/heliospheric-weather}}. An ensemble of $n$ members is used as a CME input, where the variation in model parameters is also taken into account by constructing the synthetic values of solar wind speed, $w$, and drag parameter, $\gamma$. $m$ synthetic values of $w$ and $\gamma$ are constructed under the assumption that the measurements are behaving as random variables and follow a normal distribution. The model input thus consists of $n\cdot m^2$ ensemble members, where $n$ is restricted by the observations and optimal $m$ was found as a compromise between the convergence of the output distribution and the computing time. The outputs are probability of arrival and distributions of CME transit times and arrival speeds, where the median is taken as the most likely value and the range is defined with $95\%$ confidence interval. Therefore, DBEM is able to provide probabilities of CME arrival at a specific target together with uncertainties in the arrival time and speed. 

We evaluated the performance of the new DBEM model on a refined sample taken from the study by \citet{mays15}, where it can also be compared to the performance of ENLIL. Depending on the number of CPUs and their speed, DBEM can perform thousand (on a single CPU) or more runs per second and is several orders of magnitude faster than ENLIL, which is its main advantage. Comparison between DBEM and ENLIL revealed that ENLIL performs slightly better than DBEM, most probably owing to the more realistic background solar wind conditions and a more suitable CME input (3D velocity). Despite small differences, the performance of the two models is similar and comparable to a ``standard" CME prediction error in transit time of $\sim10$ hours. We find that for this particular sample in both models fast CMEs are predicted to arrive earlier than observed, related to model limitations and possibly also to the overestimation of the CME initial speed for fast CMEs. Since both models show similarly good, as well as bad results, additional actions should probably be taken in order to improve the overall CME arrival forecast. A possible improvement might be to use different propagation models for different solar activity conditions. This approach, as well as the ensemble forecasting, would be analogous to the methods used in meteorology, where different models are used for different conditions since there is not a single model that is capable of forecasting the weather for all types of regions. In line with this view, the ``CME Scoreboard" website\footnote{\url{https://kauai.ccmc.gsfc.nasa.gov/CMEscoreboard/}} can serve as the platform to compare different models simulating/forecasting a variety of CME events occurring in real-time. Anyone is invited to submit their estimate of the arrival time of a recently observed CME in real-time to the CME Scoreboard. Therefore, it is suitable for model validation under real-time conditions and in addition provides an ensemble mean CME arrival time forecast from a variety of models and methods. The possible benefits from this type of approach are yet to be evaluated in future studies.

\acknowledgments

The research leading to these results has received funding from the European Union's Horizon 2020 research and innovation programme under the Marie Skodowska-Curie grant agreement No 745782. B. Vr\v{s}nak, J. \v{C}alogovi\'{c} and M. Dumbovi\'{c} acknowledge financial support by the Croatian Science Foundation under project 6212 ``Solar and Stellar Variability". M.T. acknowledges the support by the FFG/ASAP Programme under grant no. 859729 (SWAMI). A.M.V. and M.T. acknowledge support from the Austrian Science Fund (FWF): P24092-N16 and V195-N16.

\clearpage

\bibliographystyle{plainnat}
\bibliography{REFs}

\end{document}